\newif\ifpnas
\pnasfalse

\ifpnas

\documentclass[9pt,twocolumn,twoside]{pnas-new}
\templatetype{pnasresearcharticle}

\else

\documentclass{article}

\fi

\usepackage{amsmath,amsthm,amssymb}
\usepackage{stefan_tex}
\usepackage{relsize}
\usepackage{graphicx}

\usepackage{array,multirow}

\ifpnas\else
\usepackage{hyperref}
\fi

\theoremstyle{plain}
\newtheorem{proposition}{Proposition}

\newtheorem{theorem}[proposition]{Theorem}

\graphicspath{{./figures/}} 
 
 \newcommand{\beps}{\bar{\varepsilon}}
 \newcommand{\bbeta}{\bar{\beta}}
 \newcommand{\hbbeta}{\hat{\bbeta}}
 \newcommand{\bsigma}{\bar{\sigma}}

 \makeatletter
\def\@makefnmark{%
  \leavevmode
  \raise.9ex\hbox{\fontsize\sf@size\z@\normalfont\tiny\@thefnmark}}
\makeatother

\ifpnas
\else
\usepackage[hang,flushmargin]{footmisc} 
\newcommand\blfootnote[1]{%
  \begingroup
  \renewcommand\thefootnote{}\footnote{#1}%
  \addtocounter{footnote}{-1}%
  \endgroup
}
\fi

\begin{document}

\title{High-dimensional regression adjustments in randomized experiments}

\ifpnas
\else

\author{
Stefan Wager\textsuperscript{*\,\textdagger}
\and
Wenfei Du\textsuperscript{*}
\and
Jonathan Taylor\textsuperscript{*}
\and
Robert Tibshirani\textsuperscript{*\,\textsection}
}

\date{
\textsuperscript{*}Department of Statistics, Stanford University \\
\textsuperscript{\textdagger}Stanford Graduate School of Business \\ 
\textsuperscript{\textsection}Department of Biomedical Data Science, Stanford University
}

\maketitle

\fi

\begin{abstract}
We study the problem of treatment effect estimation in randomized
experiments with high-dimensional covariate information, and
show that essentially any risk-consistent regression adjustment
can be used to obtain efficient estimates of the average treatment effect.
Our results considerably extend the range of settings where high-dimensional regression adjustments	
are guaranteed to provide valid inference about the population average
treatment effect.
We then propose cross-estimation, a simple method for obtaining
finite-sample-unbiased treatment effect estimates that leverages 
high-dimensional regression adjustments.
Our method can be used when the regression model is estimated
using the lasso, the elastic net, subset selection, etc.
Finally, we extend our analysis to allow for adaptive specification search
via cross-validation, and flexible non-parametric regression adjustments
with machine learning methods such as random forests or neural networks.
\end{abstract}

\ifpnas
\else

\section{Introduction}

\fi

\ifpnas
\else
Randomized\blfootnote{
{\bf Significance Statement.}
As datasets get larger and more complex, there is a growing interest in using machine learning methods to enhance scientific analysis. In many settings, considerable work is required to make standard machine learning methods useful for specific scientific applications. We find, however, that in the case of treatment effect estimation with randomized experiments, regression adjustments via machine learning methods designed to minimize test set error directly induce efficient estimates of the average treatment effect. Thus, machine learning methods can be used out-of-the-box for this task, without any special-case adjustments.}
\fi
controlled trials are often considered the gold standard for estimating the effect of an intervention, as they allow for simple model-free inference about the average treatment effect on the sampled population. Under mild conditions, the mean observed outcome in the treated sample minus the mean observed outcome in the control sample is a consistent and unbiased estimator for the population average treatment effect.

However, the fact that model-free inference is possible in randomized controlled trials does not mean that it is always optimal: as argued by Fisher \cite{fisher1932}, if we have access to auxiliary features that are related to our outcome of interest via a linear model, then controlling for these features using ordinary least squares will reduce the variance of the estimated average treatment effect without inducing any bias. This line of research has been thoroughly explored:
under low-dimensional asymptotics where the problem specification remains fixed while the number of samples grows to infinity, it is now well-established that regression adjustments are always asymptotically helpful---even in misspecified models---provided we add full treatment-by-covariate interactions to the regression design and use robust standard errors
\cite{athey2016econometrics,berk2014covariance,ding2016decomposing,Fr2008,freedman2008regression,
imbens2009recent,lin2013agnostic,rosenbaum2002covariance,cochran1977}.

The characteristics of high-dimensional regression adjustments are less well understood. In a recent advance, Bloniarz et al.\! \cite{blon2016} show that regression adjustments are at least sometimes helpful in high dimensions: given an ``ultra-sparsity'' assumption from the high-dimensional inference literature, they establish that regression adjustments using the lasso \cite{bp,Ti96} are more efficient than model-free inference. This result, however, leaves a substantial gap between the low-dimensional regime---where regression adjustments are always asymptotically helpful---and the high-dimensional regime where we only have special-case results.

In this paper, we show that high-dimensional regression adjustments to randomized controlled trials work under much greater generality than previously known. We find that any regression adjustment with a free intercept yields unbiased estimates of the treatment effect. This result is agnostic as to whether the regression model was obtained using the lasso, the elastic net \cite{ZH2005}, subset selection, or any other method that satisfies this criterion.  We also propose a simple procedure for building practical confidence intervals for the average treatment effect.

Furthermore, we show that the precision of the treatment effect estimates obtained by such regression adjustments depends only on the prediction risk of the fitted regression adjustment. In particular, any risk-consistent regression adjustment can be made to yield efficient estimates of the average treatment effect in the sense of  \cite{bickel1998,hahn1998role,imbens2004nonparametric,robins1995semiparametric}.
Thus, when choosing which regression adjustment to use, practitioners are justified in using standard model selection tools that aim to control prediction error, e.g., Mallow's Cp or cross-validation.

This finding presents a striking contrast to the theory of high-dimensional regression adjustments in observational studies. In a setting where treatment propensity may depend on covariates, simply fitting low-risk regression models to the treatment and control samples via cross-validation is not advised, as there exist regression adjustments that have low predictive error but yield severely biased estimates of the average treatment effect \cite{athey2016efficient,belloni2013program,belloni2014inference,farrell2015robust}. Instead, special-case procedures are needed: For example, Belloni et al.\! \cite{belloni2014inference} advocate a form of augmented model selection that protects against bias at the cost of worsening the predictive performance of the regression model.
The tasks of fitting good high-dimensional regression adjustments to randomized versus observational data thus present qualitatively different challenges.

The first half of this paper
develops a theory of regularized regression adjustments with
high-dimensional Gaussian designs. This analysis enables us to
highlight the connection between the predictive accuracy of the regression
adjustment and the precision of the resulting treatment effect estimate, and
also to considerably improve on theoretical guarantees available in
prior work. In the second half of the paper, we build on these insights to
develop cross-estimation, a practical method for inference about average
treatment effects that can be paired with either high-dimensional regularized
regression or non-parametric machine learning methods.

\section{Setting and notation}

We frame our analysis in terms of the Neyman--Rubin potential outcomes model \cite{neyman1990application,rubin1974estimating}. 
Given $n$ i.i.d. observations $(X_i, \, Y_i, \, W_i)$, 
$ i=1, \, 2, \, \ldots, \ n$, we posit potential outcomes $Y_i^{(1)}$ and $Y_i^{(0)}$; then,
the outcome that we the actually observe is \smash{$Y_i = Y_i^{(W_i)}$}.
We focus on randomized controlled trials, where $W_i$ is independent of all pre-treatment characteristics,
\begin{equation}
\label{eq:rct}
\cb{X_i, \, Y_i^{(0)}, \, Y_i^{(1)}} \ \mathlarger{\indep} \ W_i.
\end{equation}
We take the predictors to be generated as
$X_i \sim F(\cdot) \in \RR^p,$
and assume a homoskedastic linear model in each arm,
\begin{equation} 
\label{eq:linmodel}
Y_i = c^{(W_i)}+X_i \cdot \beta^{(W_i)} + \varepsilon_i^{(W_i)}, \ c^{(w)} \in \RR, \ \beta^{(w)} \in
\RR^p,
\end{equation}
for $w = 0, \, 1$,
where \smash{$\varepsilon_i^{(W_i)}$} is mean-zero noise with variance $\sigma^2$;
more general models will be considered later.
We use the notation $n_0= \abs{\cb{i : W_i = 0}}$ and $n_1= \abs{\cb{i : W_i =1}}$.
We study inference about the average treatment effect
$\tau =\EE{Y(1) - Y(0)}$.
In our analysis, it is sometimes also convenient to study estimation
of the conditional average treatment effect.
\begin{align}
\begin{split}
\btau &= \frac{1}{n} \sum_{i = 1}^n \EE{Y_i^{(1)} - Y_i^{(0)} \cond X_i} \\
&= \bX \cdot \ppp{\beta^{(1)} - \beta^{(0)}} + c^{(1)} - c^{(0)}.
\end{split}
\end{align}
As discussed by \cite{imbens2004nonparametric}, good estimators for
$\btau$ are generally good estimators for $\tau$, and vice-versa.
In the homogeneous treatment effects model
$Y_i= c + X_i \cdot \beta + W_i \tau +\varepsilon_i$,
$\tau$ and $\bar\tau$ coincide.

\section{Regression adjustments with Gaussian designs}

Suppose that we have obtained parameter estimates
\smash{$\hc^{(w)}$}, \smash{$\hbeta^{(w)}$}, $w\in\cb{0, \, 1}$ for the linear model \eqref{eq:linmodel}
via the lasso, the elastic net, or any other method. We then get a natural estimator
for the average treatment effect:
\begin{equation}
\label{eq:estimator}
\htau = \bX \cdot \ppp{\hbeta^{(1)} - \hbeta^{(0)}} + \hc^{(1)} - \hc^{(0)}.
\end{equation}
In the case where \smash{$\hbeta^{(w)}$} is the ordinary least squares estimator for $\htau$, the behavior
of this estimator has been carefully studied by \cite{cochran1977,lin2013agnostic}.
Our goal is to characterize its behavior for generic regression adjustments \smash{$\hbeta^{(w)}$},
all while allowing the number of predictors $p$ to be much larger than the sample size $n$.

The only assumption that we make on the estimation scheme is that it be centered:
for $w \in \cb{0, \, 1}$,
\begin{align}
\label{eq:intercept}
\begin{split}
&\bY_w = \bX_w \cdot \hbeta^{(w)} + \hc^{(w)},
\end{split}
\end{align}
i.e., that the mean of the predicted outcomes matches that of the observed outcomes;
and $\hbeta^{(w)}$ is translation invariant and only depends on
\begin{equation}
\ff_\beta = \cb{X_i - \bX_{W_i}, \, Y_i - \bY_{W_i}, \, W_i}_{i = 1}^n.
\end{equation}
Here, $\bX_w$ and $\bY_w$ denote the mean of the outcomes $Y_i$ and features $X_i$
over all observations with $W_i = w$. Algorithmically, a simple way to enforce this constraint
is to first center the training samples $X_i \rightarrow X_i - \bX_{W_i}$, $Y_i \rightarrow Y_i - \bY_{W_i}$,
run any regression method on this centered data,
and then set the intercept using \eqref{eq:intercept}; this is done by default in standard
software for regularized regression, such as \texttt{glmnet} \cite{FHT2010}. We also note
that ordinary least squares regression is always centered in this sense, even after common forms of model selection.

Now, if our regression adjustment has a well-calibrated intercept as in \eqref{eq:intercept},
then we can write \eqref{eq:estimator} as
\begin{align}
\label{eq:estimator2}
\begin{split}
\htau
&=\bX \cdot \ppp{\hbeta^{(1)} - \hbeta^{(0)}} + \ppp{\hc^{(1)} - \hc^{(0)}} \\
&= \bY_1 - \bY_0 + \ppp{\bX - \bX_1} \cdot \hbeta^{(1)} - \ppp{\bX - \bX_0} \cdot \hbeta^{(0)}.
\end{split}
\end{align}
To move forward, we focus on the case where the data-generating model for $(X_i, \, Y_i)$ is Gaussian, i.e.,
$X_i \sim \nn\ppp{m, \, \Sigma}$ for some $m \in \RR^p$ and positive-semidefinite
matrix \smash{$\Sigma \in \RR^{p \times p}$}, and \smash{$Y_i - \EE{Y_i \cond X_i, \, W_i} \sim \nn(0, \, \sigma^2)$}.
For our purpose, the key fact about Gaussian data is that the mean of
independent samples is independent of the within-sample spread, i.e.,
\begin{equation}
\label{eq:orth}
\cb{X_i - \bX_{W_i}, \, Y_i - \bY_{W_i}}_{i = 1}^n \ \mathlarger{\indep} \ \cb{\bX_0, \, \bX_1, \, \bY_0, \, \bY_1},
\end{equation}
conditionally on the treatment assignments $W_1, \, ..., \, W_n$.
Thus, because \smash{$\hbeta^{(w)}$} only depends on
the centered data $X_i - \bX_{W_i}$ and $Y_i - \bY_{W_i}$, we can derive
a simple expression for the distribution of $\htau$. 
The following is an exact finite sample result, and holds no
matter how large $p$ is relative to $n$; a key observation is
that \smash{$\bX - \bX_w$} is mean-zero by randomization of the treatment
assignment, for $w = 0, \, 1$.

\begin{proposition}
\label{prop:main}
Suppose that our regression scheme for \smash{$\hc^{(w)}$} and \smash{$\hbeta^{(w)}$} is centered,
and that our data-generating model is Gaussian as above. Then, writing
$\Norm{v}_\Sigma^2 := v^\top\Sigma \, v$ for $v \in \RR^p$,
\begin{align}
\label{eq:gauss_cond}
\begin{split}
&\htau - \btau \ \Big| \ {n_0, \, n_1, \,  \hbeta^{(0)}, \, \hbeta^{(1)}}\eqd\  \nn\ppp{0, \, A}, \\
&A = \ppp{\frac{1}{n_0} + \frac{1}{n_1}} \ppp{\sigma^2  + \Norm{\hbbeta - \bbeta}_\Sigma^2}, \\
&\bbeta = \frac{n_1 \, \beta^{(0)} + n_0 \, \beta^{(1)}}{n}, \ \ \hbbeta = \frac{n_1 \, \hbeta^{(0)} + n_0 \, \hbeta^{(1)}}{n}.
\end{split}
\end{align}
\end{proposition}

If the errors in $\hbeta^{(0)}$ and $\hbeta^{(1)}$ are roughly orthogonal, then
\begin{equation}
\Norm{\hbbeta - \bbeta}_\Sigma^2 \approx \frac{n_1^2}{n^2} \Norm{\hbeta^{(0)} - \beta^{(0)}}_\Sigma^2 + \frac{n_0^2}{n^2} \Norm{\hbeta^{(1)} - \beta^{(1)}}_\Sigma^2
\end{equation}
and, in any case, twice the right-hand side is always an upper bound for the left-hand side.
Thus, the distribution of $\htau$ effectively depends on the regression
adjustments \smash{$\hbeta^{(w)}$} only through the excess predictive error
\begin{equation*}
\Norm{\hbeta^{(w)} - \beta^{(w)}}_\Sigma^2 = \EE{\ppp{\ppp{X - m} \cdot \ppp{\hbeta^{(w)} - \beta^{(w)}}}^2 \cond \hbeta^{(w)}},
\end{equation*}
where the above expectation is taken over a test set example $X$.
This implies that, in the setting of Proposition \ref{prop:main}, the main practical concern in choosing which
regression adjustment to use is to ensure that \smash{$\hbeta^{(w)}$} has
low predictive error.

The above result is conceptually related to recent work by Berk et al.\! \cite{berk2014covariance}
(see also \cite{pitkin2013improved}),
who showed that the accuracy of low-dimensional covariate adjustments using ordinary
least-squares regression depends on the mean-squared error of the regression fit; they also
advocate using this connection to provide simple asymptotic inference about $\tau$.
Here, we showed that a similar result holds for any regression adjustment on Gaussian designs,
even in high dimensions; and in the second half of the paper we will discuss how to move beyond the Gaussian case.

\subsection{Risk consistency and the lasso}

As stated, Proposition \ref{prop:main} provides the distribution of $\htau$ conditionally on
\smash{$\hbeta^{(w)}$}, and so is not directly comparable to related results in the literature. However,
whenever \smash{$\hbeta^{(w)}$} is risk consistent in the sense that
\begin{equation}
\label{eq:risk_consistent}
R\ppp{\hbeta^{(w)}} := \Norm{\hbeta^{(w)}- \beta^{(w)}}_\Sigma^2 \rightarrow_p 0,
\end{equation}
for $w = 0, \, 1$, we can asymptotically omit the conditioning.

\begin{theorem}
\label{theo:efficient}
Suppose that, under the conditions of Proposition \ref{prop:main}, we have a sequence of problems
where \smash{$\hbeta^{(w)}$} is risk consistent \eqref{eq:risk_consistent}, and $\PP{W = 1} \rightarrow \pi$. Then,
\begin{align}
\label{eq:efficient}
\begin{split}
&\sqrt{n}\ppp{\htau - \btau} \Rightarrow \nn\ppp{0, \, \frac{\sigma^2}{\pi\ppp{1 - \pi}}},
\end{split}
\end{align}
or, in other words, $\htau$ is efficient for estimating $\btau$
\cite{bickel1998,hahn1998role,imbens2004nonparametric,robins1995semiparametric}.
\end{theorem}

In the case of the lasso, Theorem \ref{theo:efficient} lets us substantially improve
over the best existing guarantees in the literature \cite{blon2016}.
The lasso estimates \smash{$\hbeta^{(w)}$} as the minimizer
over $\beta$ of
\begin{equation}
\sum_{\cb{i : W_i = w}} \frac{1}{2} \ppp{Y_i - \bY_w - \ppp{X_i - \bX_w} \cdot \beta}^2
+ n_w \lambda \Norm{\beta}_1,
\end{equation} 
for some penalty parameter $\lambda > 0$.
Typically, the lasso is used when we believe a sparse regression adjustment to be appropriate.
In our setting, it is well known that the lasso satisfies
\smash{$R\ppp{\hbeta^{(w)}} = \oo_P\ppp{\Norm{\Sigma}_{\text{op}}^2{\Norm{\beta^{(w)}}_0 \log(p)}\,\big/\,{n_w}}$},
provided the penalty parameter $\lambda$ is well chosen and $\Sigma$ does not
allow for too much correlation between features
\cite{bickel2009simultaneous,meinshausen2009lasso}.

Thus, whenever we have a sequence of problems as in Theorem \ref{theo:efficient}
where \smash{$\beta^{(w)}$} is $k$-sparse, i.e., 
\smash{$\beta^{(w)}$} has at most $k$ non-zero entries, and
$k \log(p) \,\big/\, n \rightarrow 0$,
we find that $\htau$ is efficient in the sense of \eqref{eq:efficient}.
Note that this result is much stronger than the related result of \cite{blon2016},
which shows that lasso regression adjustments yield efficient estimators $\htau$
in an ultra-sparse regime with $k \ll \sqrt{n}/\log(p)$.

To illustrate the difference between these two results, it is well known that
if $k \ll \sqrt{n}/\log(p)$, then it is possible to do efficient inference about the
coefficients of the underlying parameter vector $\beta$
\cite{javanmard2014confidence,van2014asymptotically,zhang2014confidence},
and so the result of \cite{blon2016} is roughly in line with the rest of the
literature on high-dimensional inference.
Conversely, if we only have $k \ll n/\log(p)$, accurate inference about
the coefficients of $\beta$ is in general impossible without further conditions on the covariance
of $X$ \cite{cai2015confidence,javanmard2015biasing}.
Yet we have shown that we can still carry out efficient inference about $\tau$.
In other words, the special structure present in randomized trials means that
much more is possible than in the generic high-dimensional regression setting.

\subsection{Inconsistent regression adjustments}

Even if our regression adjustment \smash{$\hbeta^{(w)}$} is not risk consistent,
we can still use Proposition \ref{prop:main} to derive unconditional results
about $\htau$ whenever
\begin{equation}
R\ppp{\hbbeta} := \Norm{\hbbeta - \bbeta}_\Sigma^2 \rightarrow_p R_\infty.
\end{equation}
We illustrate this phenomenon in the case of ridge regression, where regression
adjustments generally reduce---but do not eliminate---excess test-set risk.
Recall that ridge regression estimates \smash{$\hbeta^{(w)}$} as the minimizer
over $\beta$ of
\begin{equation}
\sum_{\cb{i : W_i = w}} \frac{1}{2} \ppp{Y_i - \bY_w - \ppp{X_i - \bX_w} \cdot \beta}^2 + n_w \lambda \Norm{\beta}_2^2.
\end{equation} 
The following result relies on random-matrix theoretic tools for analyzing the predictive risk of
ridge regression \cite{dobriban2015high}.

\begin{theorem}
\label{theo:ridge}
Suppose we have a sequence of problems in the setting of Proposition \ref{prop:main} with $n, \, p \rightarrow \infty$ and $p/n \rightarrow \gamma \in (0, \, \infty)$, such that the spectrum of the covariance $\Sigma$ has a weak limit. Following \cite{dobriban2015high}, suppose moreover that the true parameters $\smash{\beta^{(0)}}$ and $\smash{\beta^{(1)}}$ are independently and randomly drawn from a random effects model with
\begin{equation}
\EE{\beta^{(w)}} = 0 \eqand \Var{\beta^{(w)}} = \frac{\alpha^2}{p} I_{p \times p}, \text{ with } \alpha > 0.
\end{equation}
Then, selecting \smash{$\hbeta^{(w)}$} in \eqref{eq:estimator} via ridge regression
tuned to minimize prediction error, and with $\PP{W = 1} \rightarrow \pi$, we get
$\sqrt{n} \ppp{\htau - \btau} \Rightarrow \nn\ppp{0, \, S}$,
\begin{align}
\begin{split}
&S = 2\sigma^2 + \frac{\alpha^2}{\gamma}\ppp{\frac{\pi}{v_0\big(-\frac{\gamma\sigma^2}{\alpha^2(1 - \pi)}\big)} + \frac{1 - \pi}{v_1\big(-\frac{\gamma\sigma^2}{\alpha^2\pi}\big)}},
\end{split}
\end{align}
where the $v_w(-\lambda)$ are the companion Stieltjes transforms of the limiting
empirical spectral distributions for the treated and control samples,
as defined in the proof.
\end{theorem}

To interpret the above result, we note that the quantity $v_w(-\lambda)$
can also be induced via the limit \cite{bai2010spectral,marchenko1967distribution}
\begin{equation*}
\frac{1}{n_w} \tr\bigg(\bigg(\frac{1}{n_w} \sum_{\cb{i : W_i = w}} X_i X_i^\top + \lambda I_{n_w \times n_w}\bigg)^{-1}\bigg) \rightarrow_p v_w(-\lambda), 
\text{ for } \lambda > 0.
\end{equation*}
Finally, we note that the limiting variance of $\htau - \btau$ obtained via
ridge regression above is strictly smaller than the corresponding variance
of the unadjusted estimator, which converges to
$(\sigma^2 + (\pi^2 + (1 - \pi^2))\alpha^2 \tr(\Sigma)/p)/(\pi(1 - \pi))$;
this is because optimally-tuned ridge regression strictly improves over the
``null'' model \smash{$\hbeta^{(w)} = 0$} in terms of its predictive accuracy.

\section{Practical inference with cross-estimation}

In the previous section, we found that---given Gaussianity assumptions---generic regression adjustments
yield unbiased estimates of the average treatment effect, and also that
low-risk regression adjustments lead to high-precision estimators.
Here, we seek to build on this insight, and to develop simple inferential
procedures about $\tau$ and $\btau$ that attain the above efficiency guarantees,
all while remaining robust to deviations from Gaussianity or homoskedasticity.

Our approach is built around \emph{cross-estimation}, a procedure
inspired by data splitting and the work of \cite{aronow2013class,TE2002}.
We first split our data into $K$ equally-sized
folds (e.g., $K = 5 \text{ or } 10$) and then, for each fold $k = 1, \, ..., \, K$,  we compute
\begin{align}
\label{eq:kfold}
\begin{split}
\htau^{(k)} &= \bY_1^{(k)} - \bY_0^{(k)} + \ppp{\bX^{(k)} - \bX_1^{(k)}} \cdot \hbeta^{(1, \, -k)} \\
&\ \ \ \ \ \ - \ppp{\bX^{(k)}  - \bX_0^{(k)}} \cdot \hbeta^{(0, -k)}.
\end{split}
\end{align}
Here, \smash{$\bY_1^{(k)}$}, \smash{$\bY_0^{(k)}$}, etc. are moments taken over the $k$-th
fold, while \smash{$\hbeta^{(1, \, -k)}$} and \smash{$\hbeta^{(0, \, -k)}$} are centered regression
estimators computed over the $K - 1$ other folds. We then obtain an aggregate estimate
\smash{$\htau = \sum_{k = 1}^K \htau^{(k)} \, n^{(k)}/n$},
where \smash{$n^{(k)}$} is the number of observations in the $k$-th fold.
An advantage of this construction is that an analogue to the relation \eqref{eq:orth} now
automatically holds, and thus our treatment effect estimator $\htau$ is
unbiased without assumptions.
Note that the result below references both the average treatment effect $\tau$
and the conditional average treatment effect $\btau$.

\begin{theorem}
\label{theo:CE}
Suppose that we have $n$ independent and identically distributed samples
satisfying \eqref{eq:rct}, drawn from a linear model \eqref{eq:linmodel}
where $X_i$ has finite first moments and the conditional variance of \smash{$Y_i^{(w)}$}
given $X_i$ may vary.
Then, \smash{$\EE{\htau \cond X_1, \, ..., \, X_n} = \btau$}.
If, moreover, the \smash{$\hbeta^{(w, \, -k)}$} are all
risk-consistent in the sense of \eqref{eq:risk_consistent} for $k = 1, \, ..., \, K$,
\sloppy{
and both the signals \smash{$X_i \cdot \beta^{(w)}$} residuals
\smash{$Y_i - \EE{Y_i \cond X_i, W_i = w}$} are asymptotically Gaussian
when averaged, then writing $\bsigma_w^2 = \mathbb{E}[\operatorname{Var}[Y_i^{(w)} \cond X_i]]$, we have
}
\begin{equation}
\label{eq:eff2}
\sqrt{n} \ppp{\htau - \tau} \Rightarrow \nn\ppp{0, \, \frac{\bsigma_0^2}{1 - \pi} + \frac{\bsigma_1^2}{\pi} +\Norm{\beta^{(1)} - \beta^{(0)}}_\Sigma^2}.
\end{equation}
\end{theorem}

In the homoskedatic case, i.e., when the variance of $Y^{(w)}$ conditionally on $X$ does
not depend on $X$, then the above is efficient. With heteroskedasticity, the above is no longer
efficient because we are in a linear setting and so inverse-variance weighting could improve
precision; however, \eqref{eq:eff2} can still be used as the basis for valid inference about $\tau$.

\subsection{Confidence intervals via cross-estimation}

Another advantage of cross-estimation is that it allows for moment-based
variance estimates for $\htau$. Here, we discuss practical methods for building
confidence intervals that cover the average treatment effect $\tau$.
We can verify that the variance of \smash{$\htau^{(k)}$} is \smash{$V_k$} after conditioning
on the \smash{$\hbeta^{(w, \, -k)}$} and \smash{$n_w^{(k)}$}, with
\begin{align}
\label{eq:vartau}
V_k = \sum_{w \in \cb{0, \, 1}} \frac{1}{n_w^{(k)}} \, \Var{Y^{(w)} - X \cdot \hbbeta^{(-k)} \cond \hbbeta^{(-k)}}.
\end{align}
Now, the above moments correspond to
observable quantities on the $k$-th data fold,
so we immediately obtain a moment-based plug-in
estimator \smash{$\hV_k$} for $V_k$. Finally,
we build $\alpha$-level confidence intervals for $\tau$ as
\begin{equation}
\tau \in \htau \pm z_{1 - \alpha/2} \, \hV, \ \ \hV = \sum_{k = 1}^K \ppp{\frac{n^{(k)}}{n}}^2 \, \hV_k,
\end{equation}
where $z_{1 - \alpha/2}$ is the appropriate standard Gaussian quantile. In the setting of
Theorem \ref{theo:CE}, i.e., with risk consistency and bounded second moments,
we can verify that the \smash{$\htau^{(k)}$} are asymptotically uncorrelated
and so the above confidence intervals are asymptotically exact.

\subsection{Cross-validated cross-estimation}

High-dimensional regression adjustments usually rely on a tuning
parameter that controls the amount of regularization, e.g., the
parameter $\lambda$ for the lasso and ridge regression. Although
theory provides some guidance on how to select $\lambda$, practitioners
often prefer to use computationally-intensive methods such as cross-validation.

Now, our procedure in principle already allows for cross-validation:
if we estimate $\hbeta^{(0, \, -k)}$ in \eqref{eq:kfold} via any cross-validated
regression adjustment that only relies on all but the $k$-th data folds, then
$\htau^{(k)}$ will be unbiased for $\tau$. However, this requires running the
full cross-validated algorithm $K$ times, which can be very expensive
computationally.

Here, we show how to obtain good estimates \smash{$\htau$} using only a
single round of cross-validation. First, we specify $K$ regression folds,
and for each $k \in \cb{1, \, ..., \, K}$ and $w \in \cb{0, 1}$ we compute
\smash{$\bX_{k, \, w}$} and \smash{$\bY_{k, \, w}$} as the mean
of all observations in the $k$-th
fold with $W_i = w$. Next, we center the data
such that \smash{$\tX_i = X_i - \bX_{k, \, W_i}$}
and \smash{$\tY_i = Y_i - \bY_{k, \, W_i}$} for all observations in the $k$-th fold.
Finally, we estimate \smash{$\hbeta^{(w, \, -k)}$} by running a standard
out-of-the-box cross-validated algorithm
(e.g., \texttt{cv.glmnet} for \texttt{R}) on
the \smash{$(\tX_i, \, \tY_i, \, W_i)$}-triples
with the same $K$ folds as specified before,
and then use \eqref{eq:kfold} to compute \smash{$\htau$}.

The actual estimator that we use to estimate $\hbeta^{(0)}$ and $\hbeta^{(1)}$
in our experiments is inspired by
the procedure of Imai and Ratkovic \cite{imai2013estimating}.
Our goal is to let the lasso learn shared ``main effects'' for the treatment and control
groups. To accomplish this, we first run a $2p$-dimensional lasso problem,
\begin{equation}
\label{eq:joint_lasso}
\begin{split}
\hbeta, \, \hgamma &= \argmin_{\beta, \, \gamma} \Bigg\{ \lambda \ppp{\Norm{\beta}_1 + \Norm{\gamma}_1} \\
& \ \ \ \ \ + \sum \ppp{\tY_i - \ppp{\tX_i \cdot \beta + \ppp{2 W_i - 1} \tX_i \cdot \gamma}}^2 \Bigg\},
\end{split}
\end{equation}
and then set $\hbeta^{(0)} = \hbeta - \hgamma$ and $\hbeta^{(1)} = \hbeta + \hgamma$.
We simultaneously tune $\lambda$ and estimate $\tau$ by
cross-validated cross-estimation as discussed above.
When all our data is Gaussian,
this procedure is exactly unbiased by the same argument as used in Proposition
\ref{prop:main}; and even when $X$ is not Gaussian, it appears to work well in
our experiments.

\section{Non-parametric machine learning methods}

In our discussion so far, we have focused on treatment effect estimation
using high-dimensional, linear regression adjustments, and showed how to
provide unbiased inference about $\tau$ under general conditions. Here,
we show how to extend our results about cross-estimation to general
non-parametric regression adjustments obtained using, e.g., neural
networks or random forests \cite{breiman01:_random_fores}.
We assume a setting where
$$ \EE{Y(w) \cond X = x} = \mu^{(w)}\ppp{x} $$
for some unknown regression functions \smash{$\mu^{(w)}(x)$}, and
our goal is to leverage estimates \smash{$\hmu^{(w)}(x)$} obtained
using any machine learning method to improve the precision
of $\htau$, as follows:\footnote{We
note that \eqref{eq:ml_adjust} only depends on
\smash{$\hmu^{(0, \, -i)}\ppp{X_i}$} and \smash{$\hmu^{(1, \, -i)}\ppp{X_i}$}
implicitly through
\smash{$\bar{\hmu}^{(-i)}\ppp{X_i} = {n_1}/{n} \ \hmu^{(0, \, -i)}\ppp{X_i} + {n_0}/{n} \ \hmu^{(1, \, -i)}\ppp{X_i}$}.
It may thus also be interesting to estimate \smash{$\bar{\hmu}^{(-i)}\ppp{X_i}$} directly using, e.g., the ``tyranny of the
minority'' scheme of Lin \cite{lin2013agnostic}.}\textsuperscript{,}\footnote{A related
estimator is studied by Rothe \cite{rothe2016value} in the context of classical non-parametric regression
adjustments, e.g., local regression, for observational studies with known treatment propensities.}
\begin{align}
\label{eq:ml_adjust}
&\htau = \frac{1}{n} \sum_{i = 1}^n \ppp{\hmu^{(1, \, -i)}\ppp{X_i} - \hmu^{(0, \, -i)}\ppp{X_i}} \\
\notag
&\ \ \ \ \  +  \sum_{\cb{i : W_i = 1}} \! \frac{Y_i - \hmu^{(1, \, -i)}\ppp{X_i}}{n_1}
-  \sum_{\cb{i : W_i = 0}} \! \frac{Y_i - \hmu^{(0, \, -i)}\ppp{X_i}}{n_0},
\end{align}
where \smash{$\hmu^{(w, \, -i)}$} is any estimator that does not depend on the $i$-th
training example; for random forests, we set \smash{$\hmu^{(w, \, -i)}(X_i)$} to be the
``out-of-bag'' prediction at $X_i$.
To motivate \eqref{eq:ml_adjust}, we start from \eqref{eq:estimator2}, and
expand out terms using the relation
$$ \ppp{\bX - \bX_1} \cdot \hbeta^{(1)} = \frac{1}{n} \sum_{i = 1}^n \hmu^{(1)}\ppp{X_i} - \frac{1}{n_1} \!  \sum_{\cb{i : W_i = 1}} \!\!  \hmu^{(1)}\ppp{X_i},$$
where $\hmu^{(1)}(x) = x \cdot \hbeta^{(1)} + \hc^{(1)}$, etc.
The remaining differences between \eqref{eq:ml_adjust} and \eqref{eq:estimator2}
are due to the use of out-of-bag estimation to preserve randomization
of the treatment assignment $W_i$ conditionally on the corresponding regression adjustment.
We estimate the variance of $\htau$ using the formula
\begin{equation*}
\hV = \sum_{w \in \cb{0, \, 1}} \sum_{\cb{i : W_i = w}} \frac{\ppp{Y_i - \frac{n_0}{n} \,  \hmu^{(1, \, -i)}\ppp{X_i} - \frac{n_1}{n} \,  \hmu^{(0, \, -i)}\ppp{X_i}}^2}{n_w \, (n_w - 1)}.
\end{equation*}
The following result characterizes the behavior of this estimator, under the
assumption that the estimator is ``jackknife-compatible,'' meaning that the expected
jackknife estimate of variance for \smash{$\hmu^{(w)}$} converges to 0. We
define this condition in the proof,
and verify that it holds for random forests.

\begin{theorem}
\label{theo:ml}
Suppose that \smash{$\hmu$} is jackknife-compatible.
Then, the estimator $\htau$ \eqref{eq:ml_adjust} is asymptotically unbiased,
$ \EE{\htau \cond X_1, \, ..., \, X_n} = \btau + o(1/\sqrt{n})$. Moreover, if the regression
adjustments $\hmu_w$ are risk-consistent in the sense that\footnote{With
random forests, \cite{scornet2015consistency} provide such a risk-consistency result.}
$1/n \sum_{i = 1}^n (\hmu^{(w, \, -i)}(X_i) - \mu^{(w)}(X_i))^2 \rightarrow_p 0$,
and the potential outcomes \smash{$Y_i^{(w)}$} have finite second moments,
then \smash{$\htau$} is efficient and
\smash{$(\htau - \tau) / (\hV)^{1/2}$} is asymptotically standard Gaussian.
\end{theorem}

\begin{figure}

\centering

\includegraphics[width=0.49\columnwidth]{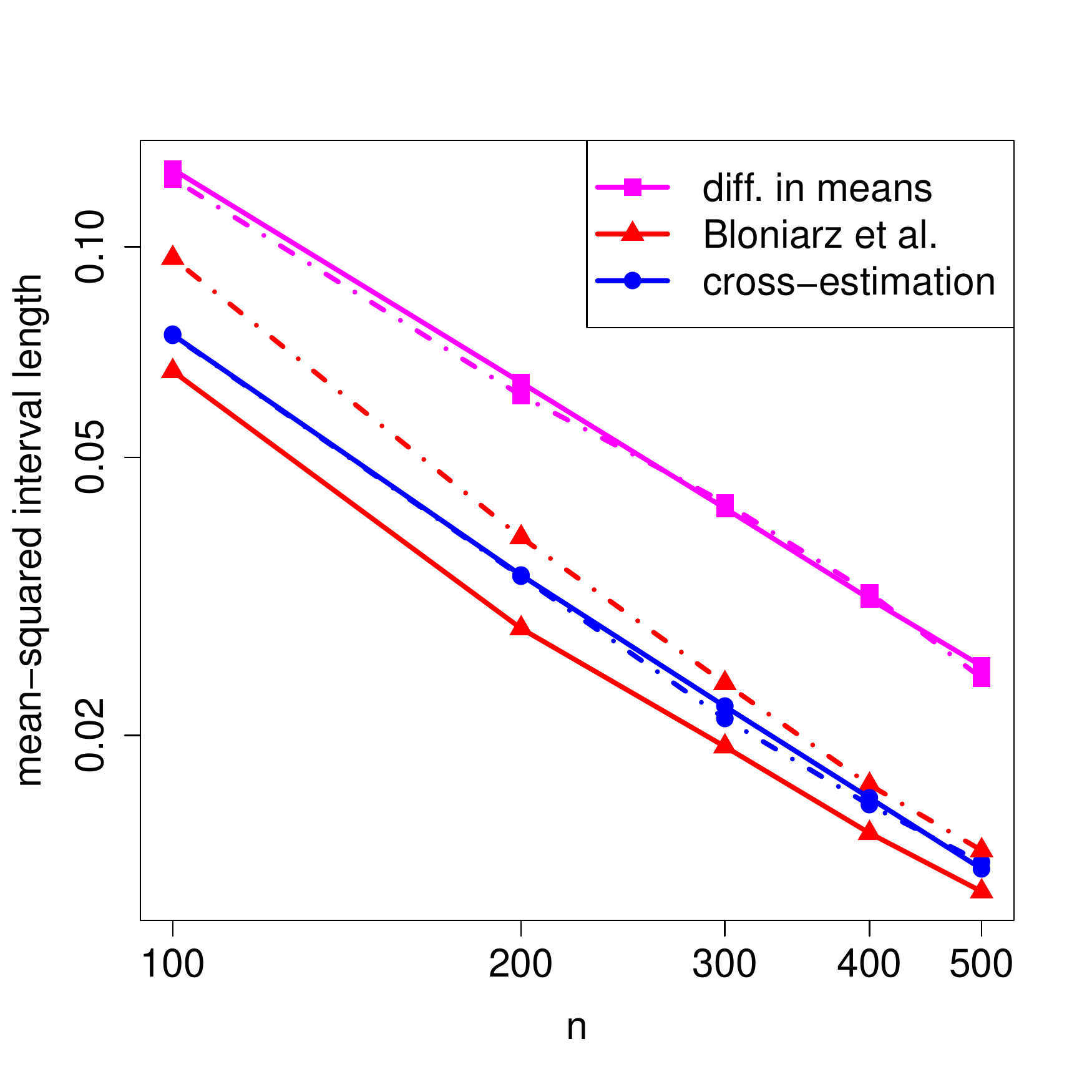}
\includegraphics[width=0.49\columnwidth]{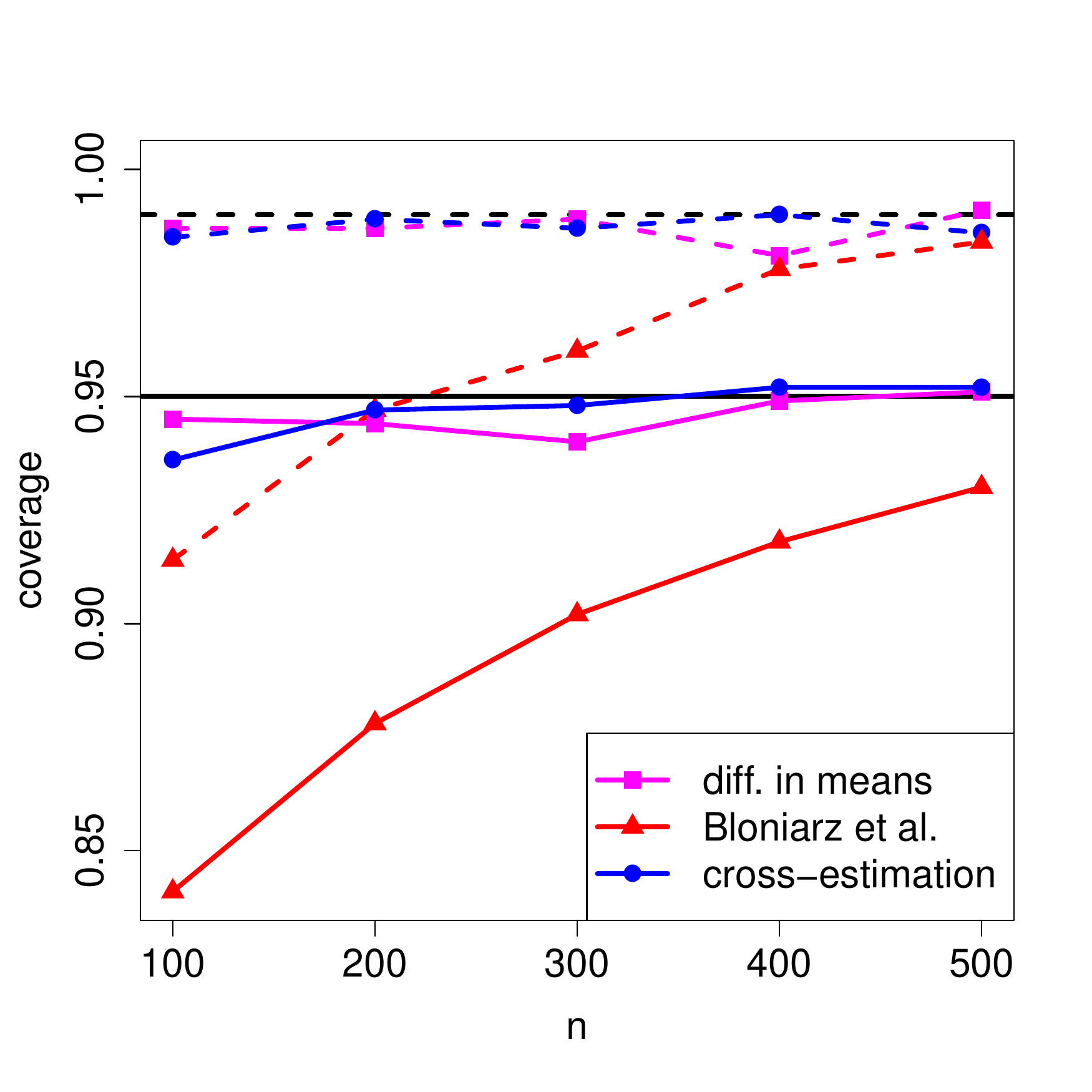}

\caption{Simulation results with $\beta = (1, \, 0, \, 0, \, ..., \, 0)$,
$\PP{W = 1} = 0.2$, $\rho = 0$, and $p = 500$. All numbers
are based on 1000 simulation replications.
The left panel shows both the average
variance estimate \smash{$\hV$} produced by each estimator (solid lines), and the
actual variance \smash{\text{Var}[$\htau$]} of the estimator (dashed-dotted lines);
note that \smash{$\hV$} is directly
proportional to the squared length of the confidence interval. The right
panels depict realized coverage for both 95\% confidence intervals (solid lines)
and 99\% confidence intervals (dashed lines).
\label{fig:simu1}}
\end{figure}

\begin{figure}

\centering

\includegraphics[width=0.49\columnwidth]{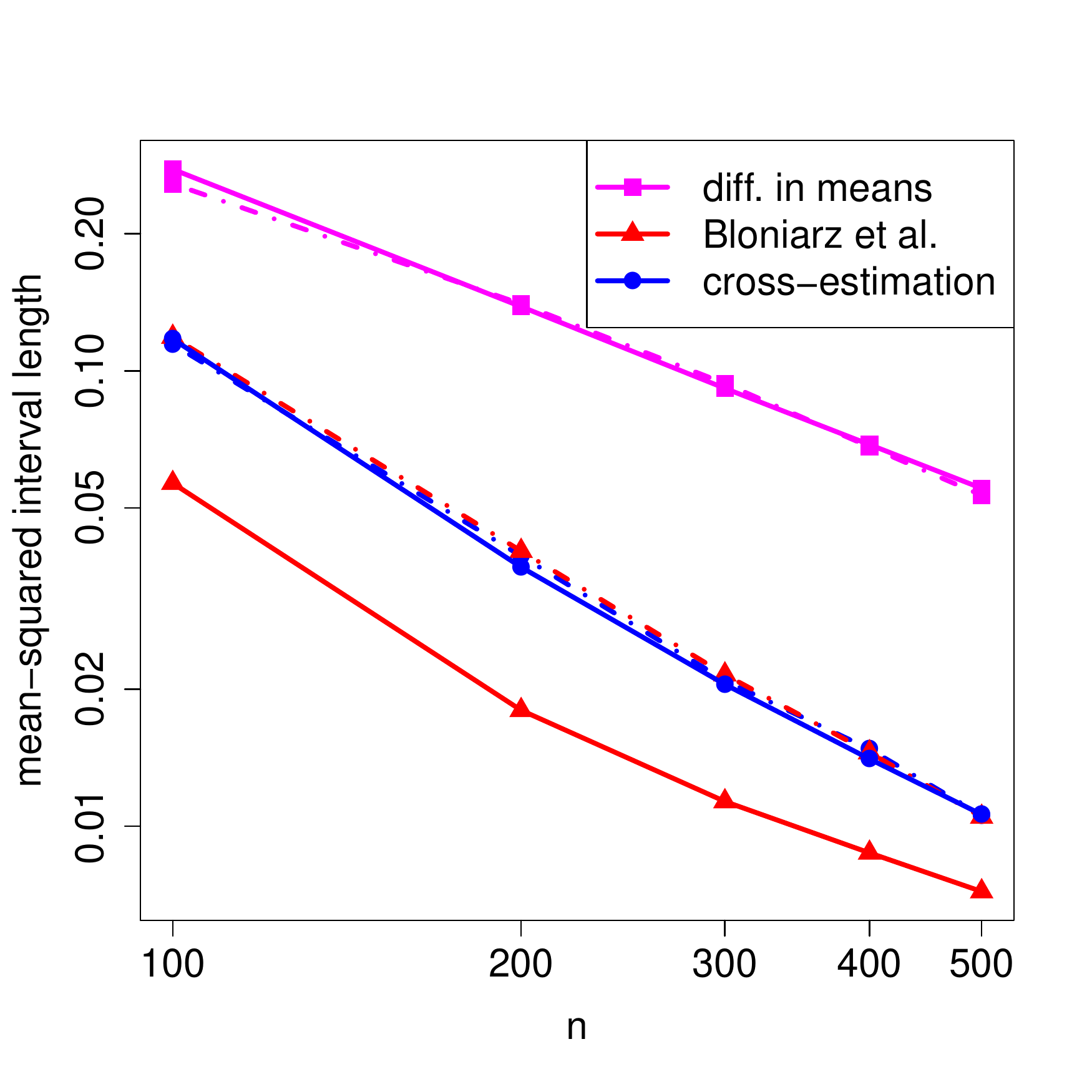}
\includegraphics[width=0.49\columnwidth]{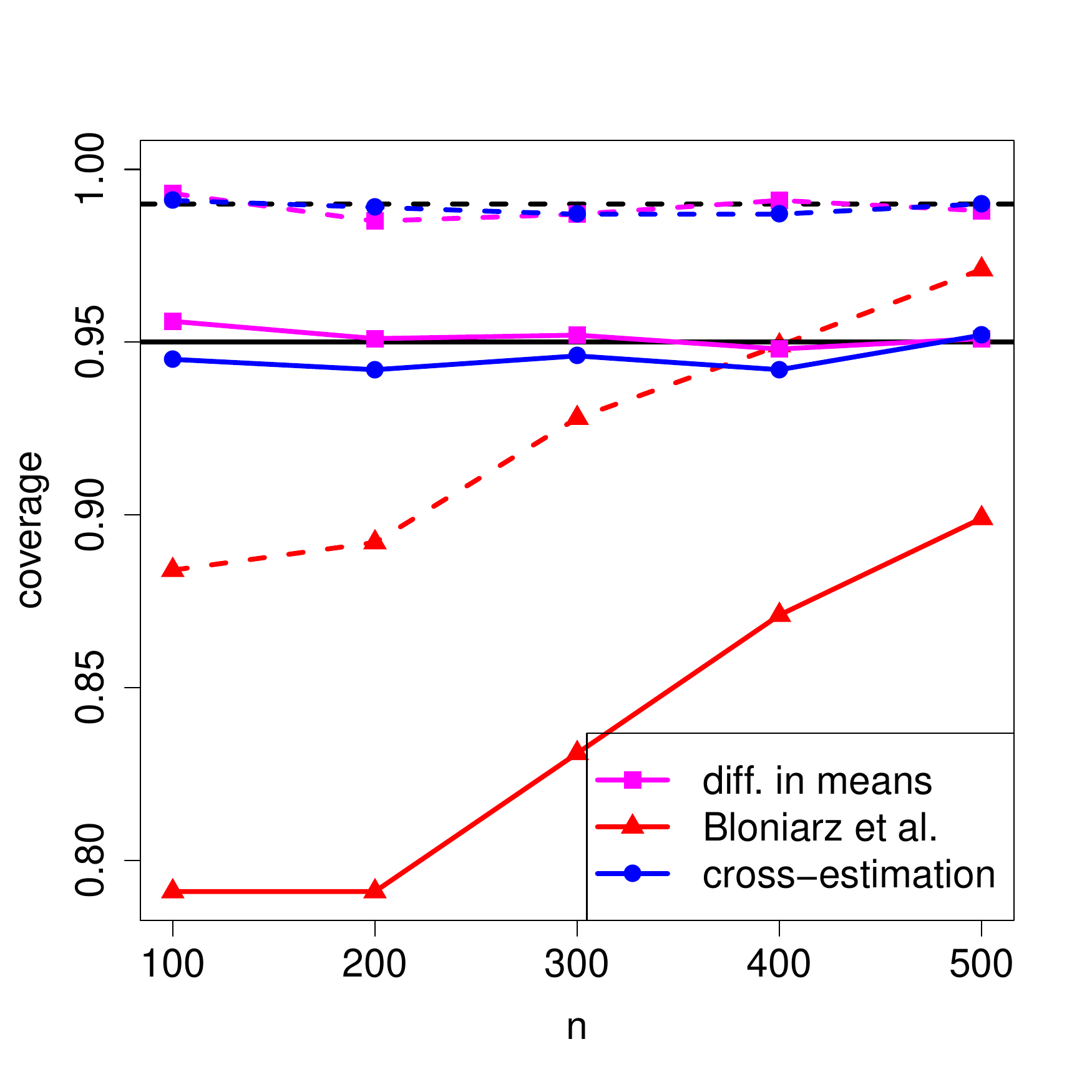}

\caption{Simulation results with $\beta$ proportional to a permutation
of $(1, \, 2^{-1}, \, 3^{-1}, \, ..., \, p^{-1})$,
$\Norm{\beta}_2 = 2$, $\PP{W = 1} = 0.5$, $\rho = 0.8$, and $p = 500$. 
All numbers are based on 1000 simulation replications.
The plots are produced the same way as in Figure \ref{fig:simu1}.\label{fig:simu2}}

\end{figure}

We note that there has been considerable recent interest in using machine
learning methods to estimate heterogeneous treatment effects
\cite{athey2016recursive,green2012modeling,hill2012bayesian,wager2015estimation}.
In relation to this literature, our present goal is more modest: we simply seek to
use machine learning to reduce the variance of treatment effect estimates
in randomized experiments. This is why we obtain more general
results than the papers on treatment heterogeneity.

\section{Experiments}

In our experiments, we focus on two specific variants of treatment
effect estimation via cross-estimation. For high-dimensional linear
estimation, we use the lasso-based method \eqref{eq:joint_lasso}
tuned by cross-validated cross-estimation.
For non-parametric estimation, we use \eqref{eq:ml_adjust} with random forest adjustments.
We implement our method as an open-source \texttt{R}-package, \texttt{crossEstimation},
built on top of \texttt{glmnet} \cite{FHT2010} and
\texttt{randomForest} \cite{liaw2002classification}
for \texttt{R}.
The supporting information has additional simulation results.

\subsection{Simulations}

We begin by validating our method in a simple simulation
setting with $Y = X\beta + W\tau + \varepsilon$, where
$\varepsilon \sim \nn(0, \, 1)$.
In all simulations, we set the the features $X$ to be Gaussian
with auto-regressive AR-$\rho$ covariance.
We compare our lasso-based
cross-estimation with both the simple difference-in-means estimate
$\htau = \bY_1 - \bY_0$, and the proposal of Bloniarz et
al.\! \cite{blon2016} that uses lasso regression adjustments
tuned by cross-validation.
Our method differs from that of Bloniarz et al.\! in that we use
a different algorithm for confidence intervals, and also that
we use the joint lasso algorithm \eqref{eq:joint_lasso}
instead of computing separate lassos in both
treatment arms.

Figures \ref{fig:simu1} and \ref{fig:simu2} display results for different
choices of $\beta$, $\rho$, etc., while varying $n$. In both cases, we see that
the confidence intervals produced by our cross-estimation algorithm and the method
of Bloniarz et al.\! are substantially shorter than those produced by the difference in means
estimator. Moreover, our confidence intervals accurately represent the variance of
our estimator (compare solid and dashed-dotted lines in the left panels), and achieve nominal
coverage at both the 95\% and 99\% levels. Conversely, especially in small samples,
the method of Bloniarz et al.\! underestimates the variance of the method, and does not
achieve target coverage.

\subsection{Understanding attitudes towards welfare}

We also consider an experimental dataset collected as
a part of the General Social Survey.\footnote{Subjects were either asked
whether we, as a society, spend too much money on ``welfare'' or
on ``assistance to the poor.'' The questions were randomly assigned
and the treatment effect corresponds to the change in the proportion of
people who answer ``yes'' to the question. This dataset is discussed
in detail in \cite{green2012modeling};
we pre-process the data as in \cite{wager_thesis}.}
The dataset is large ($N=28646$ after pre-processing), so we know the true treatment effect
essentially without error: The fraction of respondents who say we spend too
much on assistance to the poor is smaller than the fraction of respondents
who say we spend too much on welfare by 0.35. To test our method, we repeatedly
drew subsamples of size $n=2000$ from the full dataset, and 
examined the ability of both lasso- and random-forest-based
cross-estimation to recover the correct answer.
We had $p = 12$ regressors.

First of all, we note that both variants of cross-estimation
achieved excellent coverage. Given a 
nominal coverage rate of 95\%, the simple difference-in-means estimator, 
lasso-based cross-estimation and random forest cross-estimation had realized
coverage rates of 96.3\%, 96.5\% and 95.3\% respectively over 1,000 replications.
Meanwhile, given a nominal target of 99\%, the realized numbers became
99.0\%, 99.0\%, and 99.3\%. We note that this dataset has non-Gaussian features and exhibits
considerable treatment effect heterogeneity.

\begin{figure}[t]

\centering

\includegraphics[width=0.7\columnwidth]{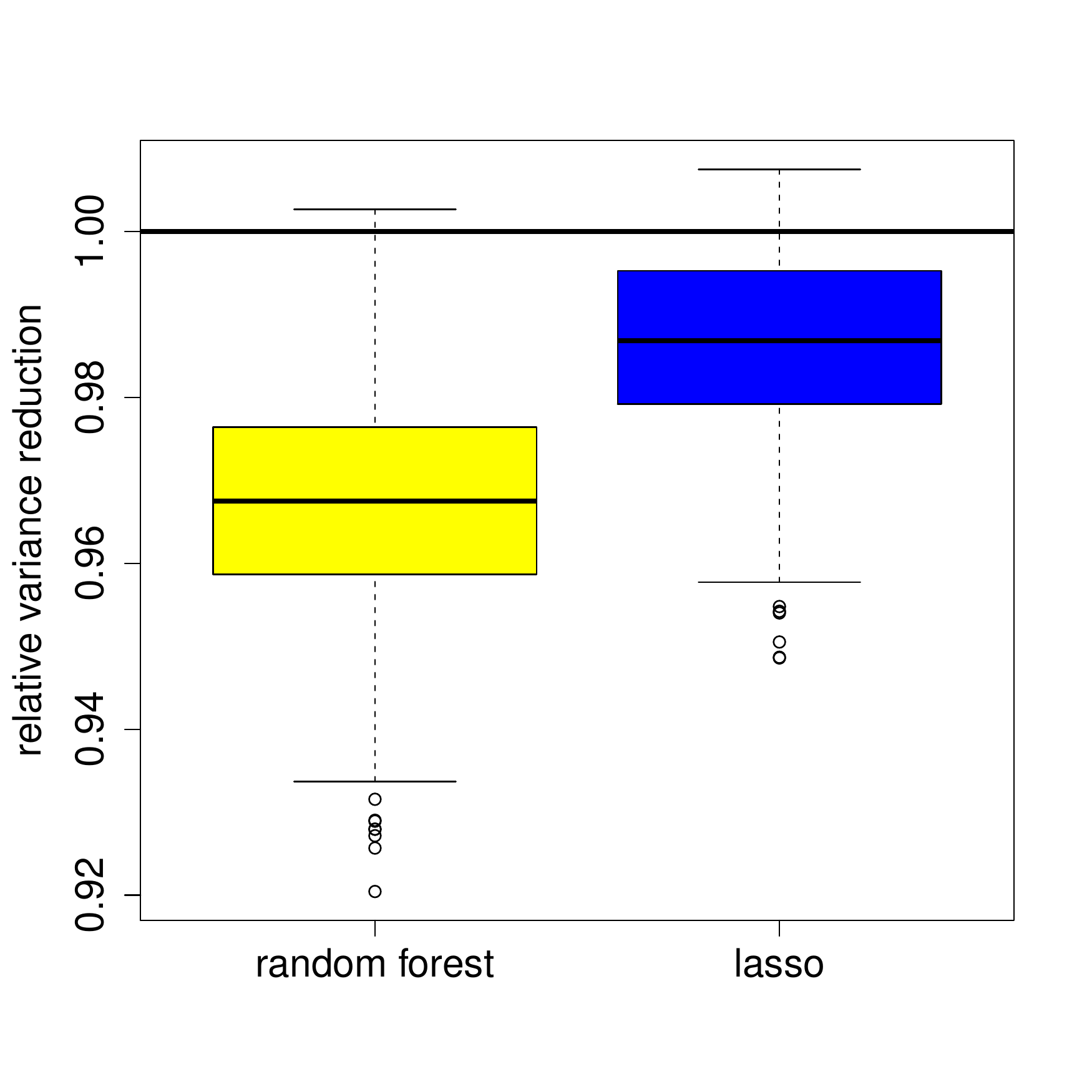}

\caption{Reduction in squared confidence interval length achieved by random forests and a lasso-based method, relative to the simple difference in means estimator. Confidence intervals rely on cross-estimation. Aggregated over 1000 replications.\label{fig:welfare}}

\end{figure}

Second, Figure \ref{fig:welfare} depicts the reduction in squared confidence
interval length for individual realizations of each method. More formally, we
show boxplots of \smash{$\hV_{\text{lasso/rf}} \, / \, \hV_{\text{simple}}$},
where \smash{$\hV$} is the variance estimate used to build confidence intervals.
Here, we see that although cross-estimation may not improve the precision
of the simple method by a large amount,  it consistently improves performance by
a small amount.  Moreover, in this example, random forests result in a larger improvement in
precision than lasso-based cross-estimation.

\vspace{-0.7\baselineskip}

\section{Discussion}

In many applications of machine learning methods to causal inference,
there is a concern that the risk of specification
search, i.e., trying out many candidate methods and choosing the one
that gives us a significant result, may reduce the credibility of
empirical findings.
This has led to considerable interests in methodologies that allow for
complex model fitting strategies that do not compromise statistical inference.

One prominent example is the design-based paradigm to causal inference
in observational studies, whereby we first seek to build an
``observational design'' by only looking at the features $X_i$ and
the treatment assignments $W_i$, and only reveal the outcomes $Y_i$
once the observational design has been set
\cite{rosenbaum2002observational,rubin2007design}. The observational
design may rely on matching, inverse-propensity weighting, or other techniques.
As the observational design is fixed before the
outcomes $Y_i$ are revealed, practitioners can devote considerable
time and creativity to fine-tuning the design without compromising
their analysis.

From this perspective, we have shown that regression adjustments to
high-dimensional randomized controlled trials exhibit a similar
opportunity for safe specification search. Concretely, imagine that
once we have collected data from a randomized experiment, we only
provide our analyst with class-wise centered data: $W_i$, $X_i - \bX_{W_i}$,
and $Y_i - \bY_{W_i}$. The analyst can then use this data to obtain any
regression adjustment they want, which we will then plug into \eqref{eq:estimator}.
Our results guarantee that---at least with a random Gaussian
design---the resulting treatment effect estimates will be unbiased
regardless of the specification search the analyst may have done
using only the class-wise centered data. Cross-estimation enables
us to mimic this phenomenon with non-Gaussian data.

\begin{appendix}

\section{Proofs}

\subsection{Proof of Proposition \ref{prop:main}}

Beginning from \eqref{eq:estimator2}, we see that
\begin{align*}
\htau - \btau
&= \bY^{(1)} - \bY^{(0)} + \ppp{\bX - \bX_1} \cdot \hbeta^{(1)} - \ppp{\bX - \bX_0} \cdot \hbeta^{(0)} \\
&\ \ \ \ \  - \ppp{c^{(1)} - c^{(0)}} - \bX \cdot \ppp{\beta^{(1)} - \beta^{(0)}} \\
&= \beps^{(1)} - \beps^{(0)} + \ppp{\bX - \bX_1} \cdot \ppp{\hbeta^{(1)} - \beta^{(1)}} \\
&\ \ \ \ \ - \ppp{\bX - \bX_0} \cdot \ppp{\hbeta^{(0)} - \beta^{(0)}} \\
&= \beps^{(1)} - \beps^{(0)} - \ppp{\bX_1 - \bX_0} \, \ppp{\hbbeta - \bbeta},
\end{align*}
where $\smash{\bbeta}$ and $\smash{\hbbeta}$ are as defined in the statement of
Proposition \ref{prop:main}, and $\beps^{(w)} = n_w^{-1} \sum_{\cb{i : W_i = w}} \varepsilon_i^{(w)}$.
Now, thanks to \eqref{eq:orth}, we can verify that that the three above summands
are independent conditionally on $n_0$, $n_1$, and $\smash{\hbbeta}$, with
\begin{align*}
&\ \beps^{(w)} \sim \nn\ppp{0, \, \sigma^2 \,/\, n_w} \eqfor w \in \cb{0, \, 1} \eqand \\
&\ppp{\bX_1 - \bX_0} \, \ppp{\hbbeta - \bbeta} \sim \nn\ppp{0, \, \ppp{\frac{1}{n_0} + \frac{1}{n_1}}  \Norm{\hbbeta - \bbeta}_\Sigma^2}.
\end{align*}

\subsection{Proof of Theorem \ref{theo:efficient}}

Given our hypotheses, we immediately see that $n_0/n \rightarrow_p 1 - \pi$,
$n_1/n \rightarrow_p \pi$, and \smash{$\lVert\hbbeta - \bbeta\rVert_\Sigma^2 \rightarrow_p 0$}.
The conclusion follows from Proposition \ref{prop:main} via Slutsky's theorem.

\subsection{Proof of Theorem \ref{theo:ridge}}

For a covariance matrix $\Sigma$, we define its spectral distribution $F(\Sigma)$
as the empirical distribution of its eigenvalues. We assume that, in our sequence of
problems, $F(\Sigma)$ converges weakly to some limiting population spectral
distribution $F_P$. Given this assumption, it is well known that the spectra
of the sample covariance matrices \smash{$\hSigma$} also converge weakly to a limiting
empirical spectral distribution $F_E$, with probability 1 \cite{bai2010spectral,marchenko1967distribution}.
In this notation, the companion Stieltjes transform $v(-\lambda)$ is defined as
$$ v(-\lambda) = \frac{1 - \gamma}{\lambda} + \gamma \, \int_0^\infty \frac{1}{z + \lambda} \ dF_E(z). $$
Given these preliminaries and under the listed hypotheses,
\cite{dobriban2015high} show that the risk of optimally tuned ridge regression
converges in probability:
$$ \frac{1}{\sigma^2} \, R\ppp{\hbeta_{\text{ridge}}} \rightarrow_p \frac{1}{\lambda^* v\ppp{-\lambda^*}} - 1,$$
where $\lambda^* = \gamma \sigma^2 / \alpha^2$ is the asymptotically optimal choice for $\lambda$.

Now, in our setting, we need to apply this result to the treatment and control samples separately.
The asymptotically optimal regularization parameters for \smash{$\hbeta^{(0)}$} and
\smash{$\hbeta^{(1)}$} are $\gamma \sigma^2 \alpha^{-2} (1 - \pi)^{-1}$ and $\gamma \sigma^2 \alpha^{-2}\pi^{-1}$
respectively. Moreover, by spherical symmetry,
$$ \ppp{\hbeta^{(1)} - \beta^{(1)}}^\top \Sigma \ppp{\hbeta^{(0)} - \beta^{(0)}} \rightarrow_p 0. $$
Thus, together we the above risk bounds, we find that
\begin{align*}
\frac{1}{\sigma^2} \Norm{\hbbeta - \bbeta}_\Sigma^2
&\rightarrow_p \pi^2 \ppp{\frac{1}{\frac{\gamma\sigma^2}{\alpha^2 (1 - \pi)}v_0\big(-\frac{\gamma\sigma^2}{\alpha^2 (1 - \pi)}\big)} - 1} \\
&\ \ \ \ \ \ \ \ + \ppp{1 - \pi}^2\ppp{\frac{1}{\frac{\gamma\sigma^2}{\alpha^2 \pi}v_1\big(-\frac{\gamma\sigma^2}{\alpha^2 \pi}\big)} - 1},
\end{align*}
where $v_0$ and $v_1$ are the companion Stieltjes transforms for the control
and treatment samples respectively.
The desired conclusion then follows from Proposition \ref{prop:main}.

\subsection{Proof of Theorem \ref{theo:CE}}

By randomization of the treatment
assignment $W_i$ we have, within the $k$-th fold and for $w \in \cb{0, \, 1}$,
\begin{align*}
&\EE{\bX_{w}^{(k)} - \bX^{(k)} \cond \cb{X_i}, \,\hbeta^{(0, \, -k)}, \, \hbeta^{(1, \, -k)}} = 0,
\eqand  \\
&\EE{\bY_1^{(k)} - \bY_0^{(k)} \cond \cb{X_i}, \, \hbeta^{(0, \, -k)}, \, \hbeta^{(1, \, -k)}} \\
&\ \ \ \ \ \ \ = c^{(1)} - c^{(0)} + \bX^{(k)} \cdot \ppp{\beta^{(1)} - \beta^{(0)}}.
\end{align*}
Thus, we see that
\begin{align*}
&\EE{\htau \cond \cb{X_i}} = c^{(1)} - c^{(0)} + \ppp{\sum_{k = 1}^K \bX^{(k)} \, \frac{n^{(k)}}{n}} \cdot \ppp{\beta^{(1)} - \beta^{(0)}} \\
&\ \ \ \ \ \ \ = c^{(1)} - c^{(0)} + \bX \cdot \ppp{\beta^{(1)} - \beta^{(0)}} = \btau. 
\end{align*}
Meanwhile, writing $\ff_k \in \cb{1, \, ..., \, n}$ for the
set of observations in the $k$-th fold and \smash{$n_w^{(k)}$} for the
number of those observations with $W_i = k$, we can write
\begin{align*}
\htau^{(k)} &=  \sum_{\cb{i \in \ff_k : W_i = 1}} \frac{Y_i - \frac{n_1^{(k)}}{n^{(k)}} \beta^{(0)} \cdot X_i - \frac{n_0^{(k)}}{n^{(k)}} \beta^{(1)}\cdot X_i}{n_1^{(k)}} \\
& \ \ -  \sum_{\cb{i \in \ff_k : W_i = 0}} \frac{Y_i - \frac{n_1^{(k)}}{n^{(k)}} \beta^{(0)}\cdot X_i - \frac{n_0^{(k)}}{n^{(k)}} \beta^{(1)}\cdot X_i}{n_0^{(k)}} \\
& \ \  + \sum_{\cb{i \in \ff_k}} \frac{(-1)^{W_i}}{n_{W_i}^{(k)}} \, \Bigg(\frac{n_1^{(k)}}{n^{(k)}} \ppp{\hbeta^{(0, \, -k)} - \beta^{(0)}} \cdot X_i \\
&\ \ \ \ \ \ \ \ \ \ \ \ \   + \frac{n_0^{(k)}}{n^{(k)}} \ppp{\hbeta^{(1, \, -k)} - \beta^{(1)}} \cdot X_i\Bigg).
\end{align*}
Now, by consistency of the regression adjustment, the third summand
decays faster that $1/\sqrt{n}$ and so can asymptotically be ignored.
Re-arranging the first two summands, we get
\begin{align*}
\htau^{(k)} &= \frac{1}{n^{(k)}} \sum_{\cb{i \in \ff_k}} \ppp{\beta^{(1)} - \beta^{(0)}} \cdot X_i \\
&\ \ \ \ \ \ \ - \sum_{\cb{i \in \ff_k}} (-1)^{W_i} \ \frac{Y_i - X_i \cdot \beta^{(W_i)}}{n^{(k)}_{W_i}} + o_P\ppp{\frac{1}{\sqrt{n}}},
\end{align*}
which has the desired asymptotic variance, and is asymptotically
Gaussian under the stated regularity conditions.

\subsection{Proof of Theorem \ref{theo:ml}}

We begin with a definition. The estimator $\hmu$ is \emph{jackknife-compatible} if,
for any  $w \in \cb{0, \, 1}$ and a new independently-drawn test point $X$,
\begin{equation}
\EE{\sum_{\cb{i : W_i = w}} \ppp{\hmu^{(w, \, -i)}(X) - \hmu^{(w)}(X)}^2 \cond n_w} \leq a(n_w)
\end{equation}
for some sequence $a(n) \rightarrow 0$. The quantity inside the left-hand expectation is the
popular jackknife estimate of variance for \smash{$\hmu^{(w)}$}; the condition requires that
this variance estimate converge to 0 in expectation. We note that this condition
is very weak: most classical statistical estimators will satisfy this condition with $a(n) = \oo(1/n)$,
while subsampled random forests of the type studied in \cite{wager2015estimation}
satisfy it with $a(n) = \oo(s/n)$, where $s$ is the subsample size.

Now, as in the proof of Theorem \ref{theo:CE}, we write $\htau$ as
\begin{align*}
&\frac{1}{n_1} \sum_{\cb{i : W_i = 1}} \ppp{Y_i - \mu^{(1)}(X_i)} - \frac{1}{n_0} \sum_{\cb{i : W_i = 0}} \ppp{Y_i - \mu^{(0)}(X_i)}\\
&\ \ \ \ \ \ + \frac{1}{n} \sum_{i = 1}^n \ppp{\mu^{(1)}(X_i) - \mu^{(0)}(X_i)} + R,
\end{align*}
where $R$ is a residual term
\begin{align*}
&R = \sum_{i = 1}^n \ \frac{(-1)^{W_i}}{n_{W_i}} \ \bigg(\frac{n_0}{n}\ppp{\hmu^{(1, \, -i)}(X_i) - \mu^{(1)}(X_i)} \\
&\ \ \ \ \ \ \ \ + \frac{n_1}{n}\ppp{\hmu^{(0, \, -i)}(X_i) - \mu^{(0)}(X_i)}\bigg).
\end{align*}
The main component of $\htau$ is an unbiased, efficient estimator for $\tau$.
It remains to show that $R$ is asymptotically unbiased given jackknife-compatibility,
and is moreover asymptotically negligible if \smash{$\hmu$} is risk-consistent.

For the remainder of the proof, we focus on the setting where the regression adjustments
\smash{$\hmu^{(0)}$} and \smash{$\hmu^{(1)}$} are computed separately on samples with
$W_i = 0$ and $W_i = 1$ respectively.
The reason $R$ may not be exactly unbiased is that $\hmu^{(1, \, -i)}(X_i)$ is a function
of $n_1$ observations if $W_i = 0$, while it is a function of $n_1 - 1$ observations if $W_i = 1$,
and this effect can create biases. Our goal is to show, however, that these biases are small for any
jackknife-compatible estimator.
To do so, we first define a ``leave-two-out'' approximation to $R$:
\begin{align*}
&R_2 = \sum_{i = 1}^n \ \frac{(-1)^{W_i}}{n_0 n_1} \!\! \sum_{\cb{j : W_j \neq W_i}} \! \bigg(\frac{n_0}{n}\ppp{\hmu^{(1, \, -\cb{i, \, j})}(X_i) - \mu^{(1)}(X_i)} \\
&\ \ \ \ \ \ \ \ + \frac{n_1}{n}\ppp{\hmu^{(0, \, -\cb{i, \, j})}(X_i) - \mu^{(0)}(X_i)}\bigg),
\end{align*}
where the \smash{$\hmu^{(w, \, -\cb{i, \, j})}(x)$} are predictions obtained without either the
$i$-th or $j$-th training examples.
We see that $\hmu^{(w, \, -\cb{i, \, j})}(X_i)$ is always computed on $n_w - 1$ observations, and
moreover, is independent of $W_i$ conditionally on $n_1$. Thus, we see that $\EE{R_2} = 0$ by
randomization, and moreover that $\EE{R_2^2} = o(1/n)$ under risk-consistency.

To establish our desired result, it remains to show that $\EE{(R_2 - R)^2} = o(1/n)$.
In the case where \smash{$\hmu^{(0)}$} and \smash{$\hmu^{(1)}$}
are computed separately on samples with
$W_i = 0$ and $W_i = 1$ respectively, we can write
\begin{align*}
&R_2 - R = \sum_{\{i : W_i = 0\}} \sum_{\{j : W_j = 1\}} \bigg(\frac{n_0}{n}\ppp{\hmu^{(1, \, -j)}(X_i) - \hmu^{(1)}(X_i)}\\
&\ \ \ \ \ \ \ \ - \frac{n_1}{n}\ppp{\hmu^{(0, \, -i)}(X_j) - \hmu^{(0)}(X_j)} \bigg) \ \bigg/ \ \ppp{n_0 n_1} .
\end{align*}
The jackknife-compatibility condition implies that
$$\EE{\ppp{\hmu^{(1, \, -j)}(X_i) - \hmu^{(1)}(X_i)}^2 \cond n_1} \leq \frac{a(n_1)}{n_1},$$
for any $j$ with $W_j = 1$, etc., and so we find that
$$ \frac{1}{2} \, \EE{\ppp{R_2 - R}^2 \cond n_1} \leq \frac{n_0^2 \, a(n_1)}{n^2 n_1} + \frac{n_1^2 \, a(n_0)}{n^2 n_0}, $$
which converges to 0 in probability at a rate faster that $1/n$, as desired.
In the case where \smash{$\hmu^{(0)}$} and \smash{$\hmu^{(1)}$}
are not computed separately, we attain the same result by applying the 
jackknife-compatibility condition on the leave-one-out estimators.

\section{Additional simulation results}

We now present additional simulation results for coverage rates of cross-validated cross-estimation,
with a focus on settings with treatment heterogeneity and potentially non-Gaussian
designs. In an effort to challenge our method,
we used signals with very high signal-to-noise ratio.
We estimate \smash{$\beta^{(0)}$} and \smash{$\beta^{(1)}$} jointly
using the procedure described in \eqref{eq:joint_lasso}. 

In these simulations, we always used an autoregressive covariance structure
with \smash{$\Sigma_{ij} = \rho^{\abs{i - j}}$}, where we interpret \smash{$0^0 = 1$}. The design
matrices $X$ were generated as \smash{$X = \Sigma^{1/2} Z$}, where $Z$ was either
Gaussian \smash{$Z_{ij} \simiid \nn(0, \, 1)$} or Bernoulli \smash{$Z_{ij} \simiid \pm 1$} uniformly
at random. The treatment assignment was random with $\PP{W_i = 1} = 0.5$.
Conditionally on $X_i$ and $W_i$, we generated
\smash{$Y_i \simiid \nn(X_i \cdot \beta^{(W_i)}, \, \sigma^2)$}.
Finally, we considered 3 different settings for the signal:
\begin{align*}
\text{Dense: }
&\beta^{(0)}_j = p^{-1}, \ \
\beta^{(1)}_j = 1.1 \, p^{-1}, \\
\text{Geometric: }
&\beta^{(0)}_j = 10^{-10 \, j/p}, \ \
\beta^{(1)}_j = 11^{-10 \, j/p}, \\
\text{Sparse: }
&\beta^{(0)}_j = 10 \cdot 1\ppp{\cb{j = 1 \text{ mod } 20}},  \\
&\beta^{(1)}_j = 9 \cdot 1\ppp{\cb{j = 1 \text{ mod } 20}} + 1\ppp{\cb{j = 1 \text{ mod } 10}},
\end{align*}
for $j = 1, \,..., \, p$.
Results are presented in Table \ref{tab:simu_more}.
Overall, the coverage rates appear quite promising, especially noting
the wide variety of simulation settings.

\begin{table}[t]
\centering
\begin{tabular}{||r|rrr||cc|cc||}
\hline
\hline
\multicolumn{4}{||r||}{$p$} & \multicolumn{2}{c}{60} & \multicolumn{2}{|c||}{500} \\
\multicolumn{4}{||r||}{$n$} & 80 & 200 & 80 & 200 \\
\hline
\hline
 \parbox[t]{1.8mm}{\multirow{8}{*}{\rotatebox[origin=c]{90}{Dense signal}}}
& $\sigma = 0.1$ & $\rho = 0$ & Gauss. $X$ & 0.94 & 0.96 & 0.95 & 0.95 \\
& $\sigma = 0.1$ & $\rho = 0$ & Bern. $X$ & 0.95 & 0.95 & 0.94 & 0.95 \\ 
& $\sigma = 0.1$ & $\rho = 0.9$ & Gauss. $X$ & 0.94 & 0.94 & 0.94 & 0.95\\ 
& $\sigma = 0.1$ & $\rho = 0.9$ & Bern. $X$ & 0.93 & 0.95 & 0.93 & 0.92 \\
\cline{2-8}
& $\sigma = 1$ & $\rho = 0$ & Gauss. $X$ & 0.95 & 0.95 & 0.92 & 0.94 \\ 
& $\sigma = 1$ & $\rho = 0$ & Bern. $X$ & 0.93 & 0.95 & 0.94 & 0.93  \\ 
& $\sigma = 1$ & $\rho = 0.9$ & Gauss. $X$ & 0.93 & 0.94 & 0.93 & 0.94  \\ 
& $\sigma = 1$ & $\rho = 0.9$ & Bern. $X$ & 0.95 & 0.96 & 0.92 & 0.94 \\
\hline \hline
\parbox[t]{2mm}{\multirow{8}{*}{\rotatebox[origin=c]{90}{Geometric signal}}}
& $\sigma = 0.1$ & $\rho = 0$ & Gauss. $X$ & 0.94 & 0.95 & 0.93 & 0.94 \\ 
& $\sigma = 0.1$ & $\rho = 0$ & Bern. $X$ & 0.92 & 0.95 & 0.94 & 0.96 \\ 
& $\sigma = 0.1$ & $\rho = 0.9$ & Gauss. $X$ & 0.95 & 0.94 & 0.95 & 0.95 \\ 
& $\sigma = 0.1$ & $\rho = 0.9$ & Bern. $X$ & 0.94 & 0.93 & 0.95 & 0.94 \\
  \cline{2-8}
& $\sigma = 1$ & $\rho = 0$ & Gauss. $X$ & 0.94 & 0.94 & 0.94 & 0.94 \\ 
& $\sigma = 1$ & $\rho = 0$ & Bern. $X$ & 0.95 & 0.94 & 0.95 & 0.95 \\ 
& $\sigma = 1$ & $\rho = 0.9$ & Gauss. $X$ & 0.91 & 0.96 & 0.94 & 0.95 \\ 
& $\sigma = 1$ & $\rho = 0.9$ & Bern. $X$ & 0.95 & 0.97 & 0.94 & 0.95 \\
  \hline \hline
 \parbox[t]{2mm}{\multirow{8}{*}{\rotatebox[origin=c]{90}{Sparse signal}}}
& $\sigma = 0.1$ & $\rho = 0$ & Gauss. $X$ & 0.94 & 0.96 & 0.94 & 0.95  \\ 
& $\sigma = 0.1$ & $\rho = 0$ & Bern. $X$ & 0.95 & 0.95 & 0.92 & 0.94 \\ 
& $\sigma = 0.1$ & $\rho = 0.9$ & Gauss. $X$ & 0.92 & 0.95 & 0.93 & 0.96  \\ 
& $\sigma = 0.1$ & $\rho = 0.9$ & Bern. $X$ & 0.93 & 0.95 & 0.92 & 0.97 \\
  \cline{2-8}
& $\sigma = 1$ & $\rho = 0$ & Gauss. $X$ & 0.94 & 0.93 & 0.94 & 0.95  \\ 
& $\sigma = 1$ & $\rho = 0$ & Bern. $X$ & 0.94 & 0.93 & 0.93 & 0.95 \\ 
& $\sigma = 1$ & $\rho = 0.9$ & Gauss. $X$ & 0.95 & 0.94 & 0.93 & 0.95  \\ 
& $\sigma = 1$ & $\rho = 0.9$ & Bern. $X$ & 0.94 & 0.94 & 0.94 & 0.94  \\
  \hline \hline
\end{tabular}
\caption{Coverage rates for 95\% nominal confidence intervals obtained by cross-validated
cross-estimation with the joint lasso procedure \eqref{eq:joint_lasso}.
All numbers are aggregated over 500 simulation runs;
the above numbers thus have a standard sampling error
of roughly 0.01.
\label{tab:simu_more}}
\end{table}

\end{appendix}

\ifpnas
\else
\bibliographystyle{abbrv}
\bibliography{tibs}
\fi

\end{document}